\documentclass[%
 reprint,
 amsmath,amssymb,
 aps,
 onecolumn
]{revtex4-2}

\usepackage{graphicx}
\usepackage{dcolumn}
\usepackage{bm}
\usepackage{color}

\begin{document}


\title{Hamiltonian formalism for  nonlinear Schr\"{o}dinger equations}

\author{Ali Pazarci}
 \affiliation{Department of Physics, Bogazici University,
34342 Bebek, Istanbul, Turkey}
\author{Umut Can Turhan}%
\affiliation{Department of Physics, Mimar Sinan Fine Arts University,
Bomonti 34380, Istanbul, Turkey}%


\author{Nader Ghazanfari}
\affiliation{Department of Physics, Mimar Sinan Fine Arts University,
Bomonti 34380, Istanbul, Turkey}%

\author{Ilmar Gahramanov}
\affiliation{Department of Physics, Bogazici University,
34342 Bebek, Istanbul, Turkey}%
\affiliation{Department of Mathematics, Khazar University,
Mehseti St. 41, AZ1096, Baku, Azerbaijan}

\date{\today}

\begin{abstract}
We study the Hamiltonian formalism for second and fourth order nonlinear Schr\"{o}dinger equations. In the case of the second order equation, we consider cubic and logarithmic nonlinearities. Since the Lagrangians generating these nonlinear equations are degenerate, we follow the Dirac-Bergmann formalism to construct their corresponding Hamiltonians. In order to obtain consistent equations of motion, the Dirac-Bergmann formalism imposes some set of constraints that contribute to the total Hamiltonian along with their Lagrange multipliers. The order of the Lagrangian degeneracy determines the number of primary constraints. If a constraint is not a constant of motion, a secondary constraint is introduced to force the consistency condition. We show that for second order and fourth order nonlinear Schr\"{o}dinger equations we only have primary constraints, and the form of nonlinearity or the order of derivatives does not change the constraint dynamics of the system. However, we observe that introducing new fields to treat higher derivatives in the Lagrangians of these equations changes the constraint dynamics, and secondary constraints are needed to construct a consistent set of Hamilton equations.        
\end{abstract}

\maketitle


\section{Introduction}
The Hamiltonian can be constructed through a straightforward Legendre transformation for a regular Lagrangian. However, in the case of the degenerate Lagrangians, for which the field variables are not independent and are connected to each other through some constraints, a new procedure is needed to construct a Hamiltonian, which produces consistent equations of motion~\cite{Sundermeyer1982}. In the late '40s and early 50's Bergmann~\cite{Bergmann1949_1, Bergman1949_2, Bergmann1950} and Dirac~\cite{Dirac1950, Dirac1951_1, Dirac1951_2} independently developed a Hamiltonian formalism for these degenerate Lagrangians, which are equivalently called singular Lagrangian systems or constrained Hamilton systems. The algorithm initially introduces a set of constraints, usually called primary constraints, with corresponding Lagrange multipliers to be determined. However, the dynamics of these constraints may also introduce the so-called secondary constraints. Such construction is important for the quantization of gauge theories in the functional integral formalism, furthermore, the constraint analysis can be used to handle higher derivative Lagrangian theories \cite{gitman2012quantization}.

Here, we study the Hamiltonian formalism of the nonlinear Schr\"{o}dinger equations (NLSEs). These equations have degenerate Lagrangians. Mathematically, a Lagrangian is called degenerate if the determinant of the Hessian matrix is zero~\cite{Sundermeyer1982}. Since there are constraints in the NLSEs caused by the singularity of the Hessian matrix, therefore, we use the Dirac-Bergmann algorithm (DBA) for constructing the consistent Hamiltonians.

NLSEs are classical field equations and are widely used to describe the properties of different physical systems. In nonlinear optics, the propagation of light in fibers and waveguides with nonlinear effects are governed by both second-order and higher-order NLSEs~\cite{Scott2004, Powers2016, Bialynicki1976}. The higher order dispersions become considerable when the pulses become extremely short~\cite{Hosseini2018}. Moreover, The Gross-Pitaevskii equation, which is a second-order NLSE, describes the general properties of trapped Bose–Einstein condensates~\cite{Pethick2008, Pitaevskii2016}. Solutions of these equations are studied in detail in the context of optics and condensed matter physics. The soliton solutions of these equations are widely investigated~\cite{novikov1984theory}, and their instabilities due to the competition between dispersion and nonlinearity are well-known~\cite{Karpman1994, Karpman1996}.

This article is organized as follows. We first describe the DBA used to construct consistent Hamilton equations of motion from a degenerate Lagrangian. Then, we apply the DBA to second-order NLSEs. We show that the Hamilton equations of motion for these systems can be constructed only by introducing primary constraints. By analyzing the already studied Korteweg-de Vries equation, we show that introducing a new field to treat the higher-order dispersion in the equation of motion generates secondary constraints. Eventually, we study the constraint dynamics of fourth-order NLSE.

\section{Dirac-Bergmann algorithm}
The DBA is a set of well-defined rules to construct the Hamiltonian from a degenerate Lagrangian~\cite{Sundermeyer1982}. For a field $\psi(\textbf{r},t)$, a Lagrangian
\begin{equation} \label{GenLag}
    L= \int d\textbf{r}\, \mathcal{L}\left[\psi_i, \nabla\psi_i, \left(\psi_i\right)_t\right],
\end{equation}
for which equations of motion are generated from the Euler-Lagrange equation
\begin{equation}
    \frac{\delta \mathcal{L}}{\delta \psi_i} - \frac{d}{dt}\frac{\delta \mathcal{L}}{\delta (\psi_i)_t} - \nabla \frac{\delta \mathcal{L}}{\delta (\nabla \psi_i)}=0,
\end{equation}
is called degenerate if the determinant of the Hessian matrix of the Lagrangian density $\mathcal{L}$ becomes zero, i.e.
\begin{equation}
    \left\vert
    \frac{\delta^2 \mathcal{L}}{\delta ({\psi_i})_t \delta ({\psi_j})_t} 
    \right\vert=0\,.
\end{equation}
Here, $\psi_t$ denotes the time derivative of the field $\psi(\textbf{r},t)$. Naturally, a Lagrangian is regular if the Hessian does not vanish. It may be seen from a Lagrangian density that a linear dependency on the total time derivatives of all fields corresponds to a degenerate Lagrangian density. 
The DBA introduces an initial set of primary constraints depending on the order of the degeneracy in the system. However, the dynamics of these constraints may also introduce a set of secondary constraints. These constraints with their corresponding multipliers contribute to the total Hamiltonian. Here, we follow the DBA step by step and construct a Hamiltonian which generates a consistent set of equations of motion.

After confirming the degeneracy of the Lagrangian, the number of the primary constraints are determined from the difference between dimension and rank of the Hessian matrix~\cite{Deriglazov2010, Lusanna2018}. These primary constraints are naturally chosen to be the canonical momenta definitions, i.e.
\begin{equation}
    \pi_{\psi_i} = \frac{\delta \mathcal{L}}{\delta (\psi_i)_t}.
\end{equation}
Each primary constraint $c_i$ contributes to the total Hamiltonian with its corresponding Lagrange multiplier $\lambda_i$. With the additional contribution of constraint $\mathcal{H}_c=\sum_i\lambda_i c_i$, the total Hamiltonian can be written as 
\begin{align}\label{defH}
     H \left[\psi_i, \pi_{\psi_i} \right] = \int d\textbf{r}\, \left(\mathcal{H}_L + \mathcal{H}_c\right),
\end{align}
where $\mathcal{H}_L = \sum_i\pi_{\psi_i} (\psi_i)_t -\mathcal{L}$ is the canonical Hamiltonian density.

The next step in the DBA is to determine the multipliers $\lambda_i$. 
Consistent equations of motion can be constructed from these constraints only if they are constant of motion, in the other words, the Poisson bracket of the constraint with the total Hamiltonian is zero, i.e. $\{c_i,H\}=0$ .
Practically, if this Poisson bracket contains any multiplier, the result can be set to zero. Accordingly, a set of vanishing Poisson brackets will give a set of linear equations for certain multipliers, which can be solved to obtain those multipliers. However, if a constraint is not a constant of motion, in other words, the Poisson bracket of the constraint with the total Hamiltonian does not contain any multiplier, the result should be forced to vanish which generates a new constraint, namely, the secondary constraint $\tilde{c}_j$. Secondary constraints are not distinct in nature from primary constraints, and their contribution to the total Hamiltonian is determined similarly but with a new set of Lagrange multipliers $\tilde{\lambda}_j$. Thus, the new constraint Hamiltonian density can be written as
\begin{equation}
    \mathcal{H}_c =\sum_{i,j} \left(\lambda_i c_i + \tilde{\lambda}_j \tilde{c}_j\right),
\end{equation}
If the consistency conditions for the new set of constraints, $\{\tilde{c}_i,H\} = 0$, are not established a new set of constraints, namely tertiary constraints, are defined and their contributions are added to $\mathcal{H}_c$ accordingly. This procedure is carried on until all multipliers are determined. Eventually, the total Hamiltonian $H$ is obtained after multiplier substitutions. Hamilton equations of motion are calculated as follows
\begin{equation}
    ({\psi_i})_t = \frac{\delta H}{\delta (\pi_{\psi_i})}, \quad
    (\pi_{\psi_i})_t = -\frac{\delta H}{\delta \psi_i}.
\end{equation}

Throughout this procedure, to construct a Hamiltonian with consistent equations of motion, we have altered the total Hamiltonian by utilizing constraints. Instead, one may alter the structure of the Poisson bracket equivalently and define Dirac brackets. However, in this manuscript, for the sake of simplicity and a better understanding of the main DBA, we are not going into the details of this procedure. In order to get a close insight into the DBA, in the following sections, we apply this procedure to different Lagrangians and construct their corresponding Hamiltonians. 
\section{Nonlinear Schr\"{o}dinger Equation}
Nonlinear Schr\"{o}dinger equations are classical field equations and are widely used to describe the properties of the different physical systems from the propagation of light in nonlinear optics~\cite{Scott2004, Powers2016} to trapped Bose–Einstein condensates~\cite{Pethick2008, Pitaevskii2016}. In this section, we will construct a Lagrangian for each of the two mostly used different equations of motion, generally called the cubic and the logarithmic nonlinear Schr\"{o}dinger equations.  

\subsection{Cubic nonlinear Schr\"{o}dinger equation}

The cubic nonlinear Schr\"{o}dinger equation
\begin{equation} \label{nls}
    i u_t + u_{xx} + 2 |u|^2 u =0 ,
\end{equation}
is one of the most studied nonlinear equations in physics~\cite{Pethick2008, Pitaevskii2016, Powers2016}. Here, $u_t$ and $u_{xx}$ are first-order time and second-order spatial derivatives of a classical field $u(x,t)$. This equation is completely integrable as a Hamiltonian system, it was solved exactly by the method of the inverse scattering by Zakharov and Shabat \cite{shabat1972exact} (see also \cite{Zakharov:1974zf,novikov1984theory}). First, we construct a Lagrangian that gives the nonlinear Schr\"{o}dinger equation as an equation of motion. Since the expression \eqref{nls} contains an imaginary part, one needs a Lagrangian with complex terms. Note that the Hamilton's equations and corresponding Poisson brackets for the nonlinear Schr\"{o}dinger equation are written in the complex form in integrability literature, see, e.g. \cite{Zakharov:1974zf}. Here, we are interested in the construction of a real-valued Lagrangian system, therefore, we factorize the nonlinear Schr\"{o}dinger equation to imaginary and real parts by setting $u(x,t)=\phi(x,t) e^{i\theta(x,t)}$ which leads us to two equations of motion
\begin{equation}\label{cubic_real_imaginary}
\begin{aligned}
  \phi_t &= -\theta_{xx}\phi-2\phi_x\theta_x,\\
  \phi\theta_t &= 2 \phi^3 +\phi_{xx}- \phi\theta_x^2 \;.
\end{aligned}
\end{equation}

The Lagrangian density whose variations with respect to $\phi$ and $\theta$ determine the $\phi_t$ and $\theta_t$ in accordance with Eq.~\eqref{cubic_real_imaginary} has the following form 
\begin{equation}\label{cnls_L}
    \mathcal{L}_{nls} \left[\phi, \phi_x, \theta_x, \theta_t \right]  = -\frac{1}{2} \theta_t \phi^2+\frac{1}{2} \phi^4 - \frac{1}{2} \phi_x^2 - \frac{1}{2} \theta_x^2 \phi^2.
\end{equation}
Since the determinant of the Hessian matrix for this Lagrangian is zero, i.e.
\begin{equation} \label{Hessian}
    \det\begin{bmatrix}
        \frac{\delta^2 \mathcal{L}}{\delta\phi_t \delta\phi_t} & \frac{\delta^2 \mathcal{L}}{\delta\phi_t \delta\theta_t}\\
        \\
        \frac{\delta^2 \delta{L}}{\delta\theta_t \delta\phi_t} & \frac{\delta^2 \mathcal{L}}{\delta\theta_t \delta\theta_t} \;.
    \end{bmatrix}=0,
\end{equation}
the Lagrangian is degenerate. The Hessian matrix has a rank of zero so the difference between the dimensionality of the matrix and its rank imposes two primary constraints. As we mentioned in the previous section, the corresponding equations for canonical momenta of fields $\theta$ and $\phi$ 
\begin{equation}\label{nls_c_momenta}
    \pi_\phi=0, \: \:\:\:\:\:\:\: \pi_\theta=-\frac{1}{2}\phi^2, \;
\end{equation}
are natural choices for the two needed primary constraints. Therefore, we set two primary constraints as
\begin{equation}\label{nls_pconstraints}
    c_1 = \pi_\phi,\:\:\:\:\:\:\:\:c_2 = \pi_\theta+\frac{1}{2}\phi^2. 
\end{equation}
By adding the contribution of the constraints $\mathcal{H}_c$ to the canonical $\mathcal{H}_L$, we can construct the total Hamiltonian density as follows 
\begin{align}\nonumber
     \mathcal{H}_{nls} & = \mathcal{H}_L + \mathcal{H}_c\\
     &= \pi_\theta \theta_t + \pi_\phi \phi_t -\mathcal{L}_{nls} + \lambda_1 c_1 + \lambda_2 c_2.
\end{align}
After the substitution of the canonical momenta and the constraints, we have
\begin{equation}
    \mathcal{H}_{nls} = -\frac{1}{2}\phi^4 +\frac{1}{2}\phi_x^2 + \frac{1}{2}\theta_x^2\phi^2 + \lambda_1 \pi_\phi + \lambda_2\left( \pi_\theta + \frac{1}{2}\phi^2 \right).
\end{equation}
In order to determine the multipliers $\lambda_1$ and $\lambda_2$, we check the so-called consistency conditions, which basically determine the preservation of the constraints under time variation. These consistency conditions, established by the Poisson brackets of the constraints with the total Hamiltonian, $H_{nls}=\int dx\, \mathcal{H}_{nls}$
\begin{equation}
    \begin{aligned}
        \{c_1,H_{nls}\} &=2\phi^3 + \phi_{xx}-\phi \theta_x^2-\lambda_2\phi,
        \\
        \{c_2,H_{nls}\} &=\lambda_1\phi + \phi^2\theta_{xx}+2\phi\phi_x\theta_x, 
\end{aligned}
\end{equation}
lead us to the determination of the multipliers $\lambda_1$ and $\lambda_2$
\begin{subequations}
    \begin{align}
        \lambda_1 &= -\theta_{xx}\phi-2\phi_x\theta_x,\\
        \lambda_2&= 2\phi^2-\theta_x^2 +\frac{\phi_{xx}}{\phi}.
    \end{align}
\end{subequations}
By substituting the determined multipliers into the total Hamiltonian density and doing the necessary cancellations, we obtain
\begin{equation}\label{integrand}
    \mathcal{H}_{nls}= \frac{1}{2}\phi^4 +\pi_\phi\left(-\phi\theta_{xx} -2\phi_x\theta_x\right) +\pi_\theta\left(2\phi^2-\theta_x^2+\frac{\phi_{xx}}{\phi}\right).
\end{equation}
Now, we can calculate the equations of motion from the total Hamiltonian
\begin{subequations}
\begin{align}\label{real_nls}
    &\phi_t =  -\theta_{xx}\phi-2\phi_x\theta_x,
    \\ \label{imaginary_nls}
    &\phi\theta_t =  2 \phi^3 +\phi_{xx} - \phi\theta_x^2,
    \\
    &\left(\pi_\phi\right)_t = 0,
    \\
    &\left(\pi_\theta \right)_t = 2\phi \phi_x \theta_x + \phi^2\theta_{xx}.
\end{align}
\end{subequations}
The first two equations \eqref{real_nls} and \eqref{imaginary_nls} are the same as the equations of motion \eqref{cubic_real_imaginary}, from which we constructed the Lagrangian \eqref{cnls_L}. If we substitute the canonical momenta into the total Hamiltonian with the integrand (\ref{integrand}) and perform integration by parts, we obtain 
\begin{equation}\label{SubstitutedNLSHamiltonian}
    H_{nls} = \int dx\, \left(  \frac{1}{2}\phi_x^2+\frac{1}{2}\phi^2\theta_x^2-\frac{1}{2}\phi^4 \right),
\end{equation}
which is a redefined version of 
\begin{equation} \label{Hnlsint}
    H_{nls} = \int dx\, \left( \frac{1}{2}|u_x|^2-\frac{1}{2}|u|^4 \right),
\end{equation}
with $u = \phi e^{i\theta}$. The expression \eqref{Hnlsint} is used in the integrability community as the Hamiltonian of the nonlinear Schrödinger equation.

\subsection{Logarithmic nonlinear Schr\"{o}dinger equation}

For the Logarithmic nonlinear Schr\"{o}dinger equation ~\cite{Bialynicki-Birula:1976tja}
\begin{equation}
    i u_t + u_{xx} + u \ln{|u|^2}=0,
\end{equation}
similar to the cubic nonlinear Schr\"{o}dinger equation, to obtain a real valued Lagrangian we set $u(x,t)=\phi(x,t) e^{i\theta(x,t)}$, which leads us to two equations of motion
\begin{equation}
 \begin{aligned}\label{log_real_imaginary}
    \phi_t &= -2\phi_x\theta_x-\phi\theta_{xx}, \\
    \phi\theta_t &= \phi_{xx} -\phi\theta_x^2+2\phi \ln\phi.
\end{aligned}   
\end{equation}
From these equations of motions, we construct the logarithmic Lagrangian density
\begin{equation}\label{Lognls_L}
    \mathcal{L}_{lnls} \left[\phi, \phi_x, \theta_x, \theta_t \right] = -\frac{1}{2}\phi^2 \theta_t -\frac{1}{2} \phi_x^2 -\frac{1}{2}\phi^2 \theta_x^2 + \phi^2 \ln{\phi} -\frac{1}{2} \phi^2.
\end{equation}
By being degenerate, having a rank zero Hessian matrix, and giving the same canonical momenta Eq.~\eqref{nls_c_momenta}, qualitatively, this logarithmic Lagrangian density is similar to the cubic one Eq.~\eqref{cnls_L}. Therefore, we choose the same primary constraints Eq.~\eqref{nls_pconstraints} and write the total Hamiltonian as follows
\begin{equation}
    \mathcal{H}_{lnls} =\frac{1}{2} \phi_x^2 + \frac{1}{2}\phi^2 \theta_x^2 +\frac{1}{2}\phi^2 -\phi^2 \ln{\phi} + \lambda_1 \pi_\phi +\lambda_2\left( \pi_\theta + \frac{1}{2} \phi^2 \right).
\end{equation}
Similarly, the two consistency conditions, defined by the Poisson brackets of the primary constraints with the total Hamiltonian, produce multipliers   
\begin{subequations}
    \begin{align}
        \lambda_1 &=  -\phi \theta_{xx} -2 \phi_x \theta_x,\\
        \lambda_2&= \frac{\phi_{xx}}{\phi} -\theta_x^2 +2\ln{\phi}.
    \end{align}
\end{subequations}
After the substitution of the multipliers, the total Hamiltonian density can be written as
\begin{equation}\label{LNLSHamiltonian}
    \mathcal{H}_{lnls}=\frac{1}{2} \phi^2 +\pi_\phi \left( -\phi \theta_{xx} -2 \phi_x \theta_x \right) + \pi_\theta \left(  \frac{\phi_{xx}}{\phi} -\theta_x^2 +2\ln{\phi} \right).
\end{equation}
And eventually, the equations of motion can be obtained as
\begin{subequations}
\begin{align}
    &\phi_t =  -2\phi_x\theta_x-\phi\theta_{xx},\\
    &\phi\theta_t =  \phi_{xx} -\phi\theta_x^2+2 \phi \ln\phi, \\
    &(\pi_\phi)_t     = 0,\\
    &(\pi_\theta)_t =  2 \phi \phi_x \theta_x + \phi^2 \theta_{xx}.
\end{align}
\end{subequations}
Similar to the case of the cubic nonlinearity, by substituting the canonical momenta and doing integration by parts the total Hamiltonian is obtained,
\begin{equation}
    H_{lnls} = \int dx\, \left(\frac{1}{2}\phi^2 +\frac{1}{2}\phi_x^2+ \frac{1}{2}\phi^2\theta_x^2-\phi^2 \ln{\phi}\right ),
\end{equation}
which can be written as
\begin{equation}
    H_{lnls} = \int dx\, \left( \frac{1}{2}|u_x|^2 +\frac{1}{2}|u|^2-\frac{1}{2}|u|^2 \ln{|u|^2} \right)
\end{equation}
with $u = \phi e^{i \theta}$.

We observe that changing the form of the nonlinearity in a degenerate Lagrangian does not generate different constraint dynamics, since its dynamics are set by the order and form of time and spatial derivatives. This result is best manifested by constraint dynamics of the Korteweg-de Vries (KdV) and the fourth-order nonlinear Schr\"{o}dinger equations. For both of these equations of motion, secondary constraints are needed to construct the Hamiltonian from a degenerate Lagrangian.     

\section{Korteweg-de Vries equation}
The KdV equation 
\begin{equation}\label{KdV_equation}
    u_t -6uu_x +u_{xxx}=0,
\end{equation}
is a useful example of a degenerate Lagrangian for which we need some secondary constraints to construct the Hamiltonian. Since studying this example helps to understand the constraint dynamics of fourth-order NLSE, we briefly review the Hamiltonian formalism of this system. The constraint dynamics of the KdV equation have already been studied in \cite{Nutku1984} (see also \cite{Nutku2001, Nutku2002}). With substitution $u(x,t)=\phi_x(x,t)$, the Lagrangian density for Eq.~\eqref{KdV_equation} can be written as
\begin{equation}\label{kdv_density}
    \mathcal{L}_{KdV}\left[\psi, \phi_x, \psi_x, \phi_t \right] =\frac{1}{2} \phi_t \phi_x +\phi_x^3+\phi_x \psi_x + \frac{1}{2}\psi^2,
\end{equation}
Here, an extra field $\psi(x,t)=\phi_{xx}(x,t)$ is introduced to avoid higher order derivatives in the Lagrangian. Although there are procedures handling the higher-order degenerate Lagrangians, introducing this extra field eases the Hamiltonian formalism of the original higher-order Lagrangian. The variations of this Lagrangian density \eqref{kdv_density} with respect to $\phi$ and $\psi$ give two equations of motion as follows
\begin{equation}
    \begin{aligned}\label{kdv_eom}
    &\phi_{xt} +6\phi_x \phi_{xx}+\psi_{xx}=0,\\
    &\psi-\phi_{xx}=0.
\end{aligned}
\end{equation}
The second equation here gives the extra defined field. Similar to the previous examples, the rank of the Hessian matrix is zero, and its difference from the dimension of the Hessian matrix imposes two primary constraints which can be chosen to be the equations for canonical momenta 
\begin{align}
    c_1 = \pi_\psi, \:\:\:\:\:\:\:\:  c_2= \pi_\phi -\frac{1}{2}\phi_x.
\end{align}
The total Hamiltonian constructed by these primary constraints is
\begin{equation}\label{arahamiltoniankdv}
    \mathcal{H}_{KdV} = -\phi_x^3 -\phi_x \psi_x -\frac{1}{2}\psi^2 +\lambda_1 c_1 + \lambda_2 c_2.
\end{equation}
Despite all similarities between the Hessian matrix properties of KdV Lagrangian and each of nonlinear Schr\"{o}dinger Lagrangians, \eqref{cnls_L} and \eqref{Lognls_L}, the consistency condition of the constraint $c_1$ does not contain any multiplier, which means we cannot obtain an expression for any multiplier. Thus, we should force the constraint to be a constant of motion. Explicitly speaking, the Poisson bracket of the constraint $c_1$ and the total Hamiltonian
\begin{equation}
    \{c_1,H\}= \psi - \phi_{xx},
\end{equation}
should be set to zero. Therefore, we need to add the secondary constraint $\tilde{c}_3 =\psi - \phi_{xx}$, along with its corresponding multiplier $\tilde{\lambda}_1$ to the total Hamiltonian density
\begin{equation}
    \mathcal{H}_{KdV} = -\phi_x^3 -\phi_x \psi_x -\frac{1}{2}\psi^2 +\lambda_1 c_1 + \lambda_2 c_2 + \tilde{\lambda}_3 \tilde{c}_3.
\end{equation}
The secondary constraint here is the result of the new field $\psi$ which is defined to treat the higher order term in the Lagrangian.
The consistency conditions lead us to the full determination of the multipliers
\begin{subequations}
    \begin{align}
        \lambda_1 &= -\phi_{5x} -6\phi_x\phi_{3x}-6\phi_{xx}^2, \\
        \lambda_2 &= -\phi_{3x}-3\phi_x^2, \\
        \tilde{\lambda}_3 &= \psi -\phi_{xx}.
    \end{align}
\end{subequations}
Accordingly, the total Hamiltonian density is obtained
\begin{equation}
    \mathcal{H}_{KdV} = \frac{1}{2}\phi_x^3 +\frac{1}{2}\psi^2 +\phi_x \psi_x +\frac{1}{2}\phi_{xx}^2- \pi_\phi\left(\phi_{3x} +3\phi_x^2\right)-\pi_\psi \left( \phi_{5x}+6\phi_{xx}^2+6\phi_x\phi_{3x} \right),
\end{equation}
and the equations of motion are calculated as follows,
\begin{subequations}
\begin{align}
    & \phi_t +3\phi_x^2 +\phi_{3x}=0,
    \\
    & \phi_{xt} +6\phi_x\phi_{xx}+\phi_{4x}=0,
    \\
    & \psi_t +6\phi_{xx}^2 +6\phi_x\phi_{3x}+\phi_{5x}=0,
    \\
    & \psi -\phi_{xx}=0.
\end{align}
\end{subequations}
Here, the second equation is the spatial derivative of the first equation, which is consistent with the third equation by substituting the fourth equation.
Note that the secondary constraint $\tilde{c}_3$ is the result of introducing the new filed $\psi$ to avoid the higher order derivatives in the Lagrangian. The constraint here generates the equation through which the field has been introduced. 

\section{Fourth-order nonlinear Schr\"{o}dinger equation}
The fourth order nonlinear Schr\"{o}dinger equation
\begin{equation}\label{fnls_eq}
    i u_t + u_{xx} + u_{4x}+ 2 |u|^2 u =0.
\end{equation}
describes the propagation of ultrashort pulses in nonlinear optics~\cite{Hosseini2018}. As in the case of second order NLSEs, to write a real-valued Lagrangian for this equation, we set $u(x,t)= \phi(x,t) e^{i\theta(x,t)}$, which leads us to the following equations of motion
\begin{equation}\label{fnls_eom}
    \begin{aligned}
        &\phi_t = -2\theta_x\phi_x +4\theta_x^3\phi_x -\phi \theta_{xx} + 6 \phi \theta_x^2 \theta_{xx} -6\theta_{xx}\phi_{xx} -4\phi_x\theta_{3x}-4\theta_{x}\phi_{3x}-\phi\theta_{4x}, \\
        &\phi\theta_t = 2 \phi^3 -\phi\theta_x^2+\phi\theta_x^4 - 12  \phi_x\theta_x\theta_{xx}-3\phi\theta_{xx}^2 +\phi_{xx}- 6\phi_{xx}\theta_x^2 -4\phi\theta_x\theta_{3x}+\phi_{4x}. 
    \end{aligned}
   \end{equation}
We write down a Lagrangian density
\begin{equation}\label{fnls_L}
    \mathcal{L}_{fnls}\left[\phi, \phi_x, \theta_x, \phi_{xx}, \theta_{xx}, \theta_{3x}, \theta_t \right] = -\frac{1}{2}\theta_t \phi^2 +\frac{1}{2}\phi^4-\frac{1}{2}\phi_x^2-\frac{1}{2}\theta
    _x^2 \phi^2 +\frac{1}{2}\theta_x^4 \phi^2  +3\theta_x^2 \phi_x^2 -2 \phi^2 \theta_x \theta_{3x}-\frac{3}{2} \theta_{xx}^2\phi^2+\frac{1}{2}\phi_{xx}^2,
\end{equation}
whose variation with respect to $\phi$ and $\theta$ leads us to the equations of motion given by \eqref{fnls_eom}. 
The dynamics of this singular Lagrangian give three primary constraints. Similar to the case of KdV, we write down a Lagrangian with a newly introduced field $\gamma=\theta_{xx}$, for which the variations with respect to fields $\phi$, $\theta$, and  $\gamma$ generate Eqs.\eqref{fnls_eom} along with $\phi^2(\gamma - \theta_{xx})=0$. Accordingly, the Lagrangian density can be written as
\begin{equation}\label{new_fnls_L}
    {\mathcal{L}_{fnls}= -\frac{1}{2} \theta _{t} \phi ^2 +\frac{1}{2} \phi ^4} -\frac{1}{2} \phi _x^2 -\frac{1}{2} \theta _x^2 \phi ^2 +\frac{1}{2} \theta _x^4 \phi^2 +\theta _x^2 \phi _x^2 -\gamma _x \theta _x \phi ^2-2 \gamma  \theta _x \phi _x \phi -\frac{1}{2} \gamma ^2 \phi ^2-2 \theta _x^2 \phi _{xx} \phi +\frac{1}{2} \phi _{xx}^2.
\end{equation}
The rank of the Hessian matrix generated from this degenerate Lagrangian density is again zero. Thus, we need three primary constraints. Note that here the dimension of the Hessian matrix is three. We set the equations for the canonical momenta as primary constraints, such that
\begin{equation}
    {c_1= \pi_\phi,
    \:\:\:\:\:\:\:\:
    c_2= \pi_\theta + \frac{1}{2}\phi^2,
    \:\:\:\:\:\:\:\:
    c_3= \pi_\gamma.}
\end{equation}
Following the DBA, we check the consistency conditions to cope with the uncertainty in the Hamilton equations. The Poisson bracket of the constraints $c_1$ and $c_2$ with the total Hamiltonian $H_{fnls}=\int dx\,\mathcal{H}_{fnls}$ give the Lagrange multipliers $\lambda_2$ and $\lambda_1$, respectively. However, The consistency condition of the constraint $c_3$ does not contain any multiplier, which means that the result should be set as a new constraint. In other words the Poisson bracket of the constraint $c_3$ and the total Hamiltonian
\begin{equation}
    \{c_3,H\}= \phi^2(\gamma-\theta_{xx}).
\end{equation}
produces a secondary constraint. We define this secondary constraint as $\tilde{c}_4=\gamma-\theta_{xx}$. By adding the contribution of the primary and secondary constraints, the total Hamiltonian density becomes
\begin{align}\nonumber
    {\mathcal{H}_{fnls}}&= \gamma _{x} \theta _{x} \phi ^2 + 2 \gamma  \theta _{x} \phi _{x} \phi +\frac{1}{2} \gamma ^2 \phi ^2 +2 \theta _{x}^2 \phi _{xx} \phi -\theta _{x}^2 \phi _{x}^2-\frac{1}{2} \theta _{x}^4 \phi ^2+\frac{1}{2} \theta _{x}^2 \phi ^2+\frac{1}{2} \phi _{x}^2-\frac{1}{2} \phi _{xx}^2-\frac{1}{2} \phi ^4 \\ 
    & \quad  
    +\lambda_1 \pi_\phi +\lambda_2 \left(\pi_\theta +\frac{1}{2} \phi ^2\right)+\lambda_3 \pi_\gamma + \tilde{\lambda}_4 \left( \gamma-\theta_{xx}\right).
\end{align}
Since there are no tertiary constraints for Lagrangian \eqref{new_fnls_L}, we are able to obtain all the multipliers from consistency conditions, as follows
\begin{subequations}
\begin{align}
    \lambda_1 &= -2 \left[2 \gamma _x \phi _x+\phi _{xx} \left(\gamma +2 \theta _{xx}\right)-2 \theta _x^3 \phi _x
     +\theta _x \left(\phi _x+2 \phi _{3x}\right)\right]-\phi  \left[\gamma _{xx}+\left(1-6 \theta _x^2\right) \theta _{xx}\right]
    +\frac{2 \phi _x^2}{\phi}\left(\theta _{xx}-\gamma \right),
    \\
    \lambda_2 & = \frac{1}{\phi }\left[ 2 \gamma  \theta _{xx} \phi -\gamma ^2 \phi -12 \theta _x \theta _{xx} \phi _x-6 \theta _x^2 \phi _{xx}
    +\left(\theta _x^4-\theta _x^2-4 \theta _{3x} \theta _x-4 \theta _{xx}^2\right) \phi +\phi _{xx}+\phi _{4x}+2 \phi ^3\right],
    \\
    \nonumber
    \lambda_3 &= -2 \left(-\gamma _{xx} \theta _{xx}-2 \gamma _{x} \theta _{3x}+\gamma _{x}^2+\gamma  \gamma _{xx}-\gamma  \theta _{4x}-2 \theta _{3x} \theta _{x}^3-6 \theta _{xx}^2 \theta _{x}^2+\theta _{3x} \theta _{x} +2 \theta _{5x} \theta _{x}+\theta _{xx}^2+6 \theta _{3x}^2
    +8 \theta _{xx} \theta _{4x}-2 \phi _{x}^2\right)  \\\nonumber
    & \quad +\frac{1}{\phi^2}\big\{ 24 \left(\theta _{xx}^2+\theta _{x} \theta _{3x}\right) \phi _{x}^2 +2 \phi _{x} \left(6 \theta _{x}^2 \phi _{3x}+30 \theta _{xx} \theta _{x} \phi _{xx}-\phi _{3x}-\phi _{5x}\right)+\phi _{xx} \left[\left(6 \theta _{x}^2-1\right) \phi _{xx}-\phi _{4x}\right]  \big\} \\
    & \quad  + \frac{1}{\phi} \left[-6 \theta _{x}^2 \phi _{4x} -12 \theta _{x} \left(3 \theta _{3x} \phi _{xx}+\theta _{4x} \phi _{x}\right)-36 \theta _{xx}^2 \phi _{xx}-36 \theta _{xx} \left(\theta _{3x} \phi _{x}+\theta _{x} \phi _{3x}\right)+\phi _{4x}+\phi _{6x} \right] \\\nonumber
    & \quad +\frac{2 \phi _{x}^2}{\phi ^3} \left(-6 \theta _{x}^2 \phi _{xx}-12 \theta _{xx} \theta _{x} \phi _{x}+\phi _{xx}+\phi _{4x}\right) +4 \phi  \phi _{xx},
    \\
    \tilde{\lambda}_4 &= \phi ^2 \left(\theta _{xx}-\gamma \right).
\end{align}    
\end{subequations}
After the substitution of the multipliers, the total Hamiltonian can be written as
\begin{equation}
    \mathcal{H}_{fnls} = \frac{1}{2 \phi ^3} \left( a_7 \phi^7 + a_6\phi^6 + a_5 \phi^5 + a_4 \phi^4  + a_3 \phi^3 + a_2 \phi^2 + a_1\phi + a_0 \right),
\end{equation}
with coefficients given as
\begin{subequations}
\begin{align} 
        a_7 &=1, \\  
        a_6 &= 0 ,\\
        a_5 &=  4\pi_\theta  + 6 \gamma \theta _{xx} - 2\gamma ^2 -6 \theta _{xx}^2+ 2 \gamma _x \theta _x - 4 \theta _x \theta _{3x}, \\
        a_4 &= 4 \gamma \theta_x\phi_x -2 \phi _{xx} \theta _x^2-12 \phi _x \theta _{xx} \theta _x-2 \pi_\phi \left[\gamma _{xx}+\left(1-6 \theta _x^2\right) \theta _{xx}\right]
    +8 \pi_\gamma \phi _{xx}+\phi _{xx}+\phi _{4x}, \\
        a_3 & = \phi _x^2 - 2 \theta _x^2 \phi _x^2 - \phi _{xx}^2 + 2\pi_\theta \left(-\gamma ^2 +2\gamma  \theta _{xx} + \theta _x^4-\theta _x^2 - 4 \theta _{3x} \theta _x-4 \theta _{xx}^2 \right) \\ \nonumber 
        & \quad - 4 \pi_\phi \left(\gamma \phi_{xx} -2 \phi _x \theta _x^3+ \phi _x \theta _x+2 \phi _{3x}\theta _x +2 \gamma _x \phi _x+2 \theta _{xx} \phi _{xx} \right) \\\nonumber
    & \quad  - 4 \pi_\gamma \left(\gamma \gamma_{xx} - \gamma \theta_{4x} -2 \theta _{3x} \theta _x^3-6 \theta _{xx}^2 \theta _x^2+\theta _{3x} \theta _x+2 \theta _{5x} \theta _x+\gamma _x^2-2 \phi _x^2 + \theta _{xx}^2+6 \theta _{3x}^2-\gamma _{xx} \theta _{xx}-2 \gamma _x \theta _{3x}+8 \theta _{xx} \theta _{4x}\right), \\
        a_2 & = \pi_\theta \left( 12 \theta_x \phi_x \theta_{xx} - \phi_{xx} + 6\theta_x^2\phi_{xx} - \phi_{4x} \right) + 4 \pi_\phi \left( \theta _{xx} \phi _x^2 -  \gamma \phi _x^2\right)  \\ \nonumber
    & \quad + 2\pi_\gamma \left[ -6 \phi _{4x} \theta _x^2- 12 \theta _x\left(3 \phi _{xx} \theta _{3x}+\phi _x \theta _{4x}\right)-36 \theta _{xx}^2 \phi _{xx}-36 \theta _{xx} \left(\phi _x \theta _{3x}+\theta _x \phi _{3x}\right)+\phi _{4x}+\phi _{6x} \right], \\
        a_1 &= 2 \pi_\gamma \big\{ 24 \left(\theta _{xx}^2+\theta _x \theta _{3x}\right) \phi _x^2 + 2 \left(6 \phi _{3x} \theta _x^2+30 \theta _{xx} \phi _{xx} \theta _x-\phi _{3x}-\phi _{5x}\right) \phi _x+\phi _{xx} \left[ \left(6 \theta _x^2-1\right) \phi _{xx}-\phi _{4x}\right] \big\},  \\
        a_0 &= 4 \pi_\gamma \phi _x^2 \left(-6 \phi _{xx} \theta _x^2-12 \phi _x \theta _{xx} \theta _x+\phi _{xx}+\phi _{4x}\right).
\end{align}
\end{subequations}
In order to show the consistency we produce all desired equations of motion of this Hamiltonian which are obtained as  
\begin{subequations}
    \begin{align}
        &{\phi_t} = -2\theta_x\phi_x +4\theta_x^3\phi_x -\phi \theta_{xx} + 6 \phi \theta_x^2 \theta_{xx} -6\theta_{xx}\phi_{xx} -4\phi_x\theta_{3x}-4\theta_{x}\phi_{3x}-\phi\theta_{4x}, \\
        &{\phi\theta_t }= 2 \phi^3 -\phi\theta_x^2+\phi\theta_x^4 - 12  \phi_x\theta_x\theta_{xx}-3\phi\theta_{xx}^2 +\phi_{xx}- 6\phi_{xx}\theta_x^2 -4\phi\theta_x\theta_{3x}+\phi_{4x}, \\
        &{ \gamma_t}=(\theta_{t})_{xx}=(\theta_{xx})_{t}, \\
        & {(\pi_\phi)_t}=0, \\ 
        & {(\pi_\gamma)_t}=0.
    \end{align}   
\end{subequations}
By substituting the canonical momenta the total Hamiltonian can be written as
\begin{align}
     {H_{fnls}} &= \int dx\, \left(
    2 \theta _x \theta _{xx} \phi _x \phi +2 \theta _x^2 \phi _{xx} \phi -\theta _x^2 \phi _x^2-\frac{1}{2} \theta _x^4 \phi ^2+\frac{1}{2} \theta _x^2 \phi ^2+\frac{1}{2} \theta _{xx}^2 \phi ^2+\theta _x \theta _{3x} \phi ^2+\frac{1}{2} \phi _x^2-\frac{1}{2} \phi _{xx}^2-\frac{1}{2} \phi ^4 \right) \;.
\end{align}
 Eventually, the total Hamiltonian as a functional of the original field $u=\phi e^{i\theta}$ is obtained as \cite{Karpman1991}
\begin{equation}
    H_{fnls} = \int dx\, \left( \frac{1}{2}|u_x|^2 -\frac{1}{2}|u|^4-\frac{1}{2}|u_{xx}|^2 \right) \;.
\end{equation}
\section{Conclusion}
The Dirac-Bergmann algorithm is used to construct the Hamiltonian from degenerate Lagrangians. The algorithm introduces a set of constraints and their corresponding Lagrange multipliers in order to obtain consistent Hamilton equations of motion. By studying different degenerate Lagrangians, we conclude that, for nonlinear Schr\"{o}dinger equations, the form of nonlinearity or the order of spatial derivatives does not change the dynamics of the constraints. However, introducing a new field to treat the higher-order derivatives appearing in the Lagrangian generates secondary constraints. The constraints naturally are the same as the definition of the new fields.    

\bibliography{NLSE}

\end{document}